\documentclass[conference]{IEEEtran}

\usepackage[normalem]{ulem}
\usepackage{booktabs}   
\usepackage{subfig}

\usepackage{graphicx}
\usepackage{tabularx}
\usepackage{algorithm}
\usepackage[noend]{algpseudocode}
\usepackage{alltt}
\usepackage{url}
\usepackage{footnote}
\usepackage{amsmath}
\usepackage{doi}
\usepackage{xparse}
\usepackage{placeins}
\usepackage{listings}
\usepackage{color}
\definecolor{lightgray}{rgb}{.9,.9,.9}
\definecolor{darkgray}{rgb}{.4,.4,.4}
\definecolor{purple}{rgb}{0.65, 0.12, 0.82}

\begin{document}

\title{Genet: A Quickly Scalable Fat-Tree Overlay for Personal Volunteer Computing using WebRTC} 

\author{\IEEEauthorblockN{Erick Lavoie, Laurie Hendren}
\IEEEauthorblockA{\textit{School of Computer Science} \\
\textit{McGill University}\\
Montreal, Canada, \\
erick.lavoie@mail.mcgill.ca, \\ hendren@cs.mcgill.ca}
\and
\IEEEauthorblockN{Fr\'ederic Desprez}
\IEEEauthorblockA{\textit{Centre de recherche Grenoble Rh\^one-Alpes} \\
\textit{INRIA}\\
Grenoble, France, \\
frederic.desprez@inria.fr}
\and
\IEEEauthorblockN{Miguel Correia}
\IEEEauthorblockA{\textit{INESC-ID/Instituto Superior T\'ecnico} \\
\textit{Universidade de Lisboa}\\
Lisboa, Portugal, \\
miguel.p.correia@tecnico.ulisboa.pt}}

\maketitle

\begin{abstract} 
WebRTC enables browsers to exchange data directly but the number of possible concurrent connections to a single source is limited. We overcome the limitation by organizing participants in a fat-tree overlay: when the maximum number of connections of a tree node is reached, the new participants connect to the node's children. Our design \textit{quickly scales} when a large number of participants join in a short amount of time, by relying on a novel scheme that only requires \textit{local information} to route connection messages: the destination is derived from the hash value of the combined identifiers of the message's source and of the node that is holding the message. The scheme provides \textit{deterministic routing} of a sequence of connection messages from a single source and \textit{probabilistic balancing} of newer connections among the leaves. We show that this design puts at least 83\% of nodes at the same depth as a deterministic algorithm, can connect a thousand browser windows in 21-55 seconds in a local network, and can be deployed for volunteer computing to tap into 320 cores in less than 30 seconds on a local network to increase the total throughput on the Collatz application by two orders of magnitude compared to a single core.
\end{abstract}

\section{Introduction} 

Web browsers provide a fast execution environment~\cite{Khan:2014,herrera18webassembly}, are already installed on most devices today, and are progressively adding support for WebRTC~\cite{webrtc}, which enables direct connections for exchanging data.  However, current browser implementations limit the number of concurrent WebRTC connections to a single source to 256~\cite{webrtc-limit}. In practice, this number is even more limited: the overhead of maintaining connections becomes significant beyond 70 concurrent connections in some libraries.\footnote{Such as the \texttt{electron-webrtc}~\cite{electron-webrtc} library for Node.js.} This limits the total number of participants.

We increased the total number of participants by using a \textit{fat-tree overlay}. In a fat-tree~\cite{Leiserson85}, processors are located on the leaves and internal nodes relay data for all their children; each new layer in the tree increases the number of possible connections exponentially. The main benefit in the context of the Web is to remove the need to relay data on dedicated servers by employing intermediate nodes in a fat-tree as relays. 

 Existing work on fat-trees~\cite{Leiserson85,ruft-fat-tree,adaptive-vs-deterministic-routing,birrer2004fatnemo} has not so far provided solutions for quick scaling. The key issues are to quickly distribute the newer participants among the existing leaves and quickly route, through the fat-tree, the multiple messages generated by WebRTC to open a new connection. To address both, we propose a novel routing scheme, which we call Genet~\footnote{Clonal colony of plants in which all individuals share the same genetic material. We expect the design, if it is successful, to eventually form a dynamic forest of overlays on the Web.}, that only requires \textit{local information} to route connection messages: this eliminates the latency that would otherwise be incurred by waiting for the status from other parts of the tree. The destination for messages is derived from the hash value of the combined identifiers of the source and the current routing node, providing two properties. First, the scheme \textit{deterministically routes} multiple messages sent by a new participant to the same leaf node. Second, the scheme ensures  \textit{probabilistic balancing} of newer connections between all the children to keep the tree balanced. This design is especially suited to the context of compute-intensive applications that leverage volunteers' devices because users tend to add local devices first before asking for help from others; the devices in the first layer of the tree will therefore also benefit from the largest available bandwidth.

To show the \textit{probabilistic balancing} scheme is useful, we measured the depth of nodes and found that at least 83\% of nodes have the same depth as they would have in a deterministic scheme; this percentage grows to 92.5\% as the tree grows larger and is independent of the failure level of nodes if they reconnect through the root after a disconnection. To show the design can quickly scale, we measured the time required for all participants to become connected within a fat-tree overlay fully implemented and tested in Pando~\cite{lavoie2018pando}, a tool for personal volunteer computing~\cite{lavoie2018pvc} that targets shorter-running tasks than is typical for well-known and larger-scale volunteer computing projects~\cite{boincprojects}. We succeeded in connecting a thousand browser windows in 22-55 seconds on a local network and could fully deploy the Collatz application on 320 cores, reaching maximum throughput in less than 30 seconds. Both results show that the design is quite useful for quick deployments on local networks, such as those in a university department or a large organization. Additional preliminary measurements of connectivity probability and latency for WebRTC on Internet deployments show that  further refinements of the design in an Internet setting shall include tolerance to failures of initial connections, perhaps by initiating multiple connections upon joining, and tolerate initial connection latencies of up to 9-16 seconds.

Compared to previous work on fat-trees, we are the first to (1)  propose a deterministic routing scheme for connection messages to quickly grow a fat-tree overlay when a large number of participants join in a short amount of time, (2) implement such a design with WebRTC to overcome the limit on the number of connections, and (3) apply the idea to dispatch work and retrieve results in a volunteer computing tool, using participants for data distribution rather than a dedicated server. 

In the rest of this paper, we first explain the design of the fat-tree overlay in more detail (Section~\ref{Section:FatTreeOverlay}). We then explain how we adapted Pando to use our fat-tree overlay to improve its scalability (Section~\ref{Section:Application}). We continue with an evaluation of the resulting implementation (Section~\ref{Section:Evaluation}). We then review similar work (Section~\ref{Section:RelatedWork}) and summarize the contributions of the paper (Section~\ref{Section:Conclusion}).

\section{Design}
\label{Section:FatTreeOverlay}

Our \textit{fat-tree overlay} organizes participants in a tree to increase the number of concurrent connections that can be made to a single origin, while bounding the number of concurrently active WebRTC connections each participant maintains.  To establish a WebRTC connection, participants exchange \textit{signals}, or possible connection endpoints, with one another to determine how to connect through the Network-Address Translation (NAT) schemes used by routers. The ICE signalling protocol~\cite{ice-draft} used by WebRTC uses a \textit{trickle mode} in which signals are sent as they are discovered. This reduces the latency to open the connection compared to waiting for all endpoints to be identified. The trickle mode generates multiple messages that need to be routed through the tree to exactly the same destination node. Moreover, to minimize the latency and make the tree grow quickly, the depth of nodes should be minimized by making the number of children in sibling sub-trees similar. 

\begin{figure}[htbp]
\includegraphics[width=0.48\textwidth]{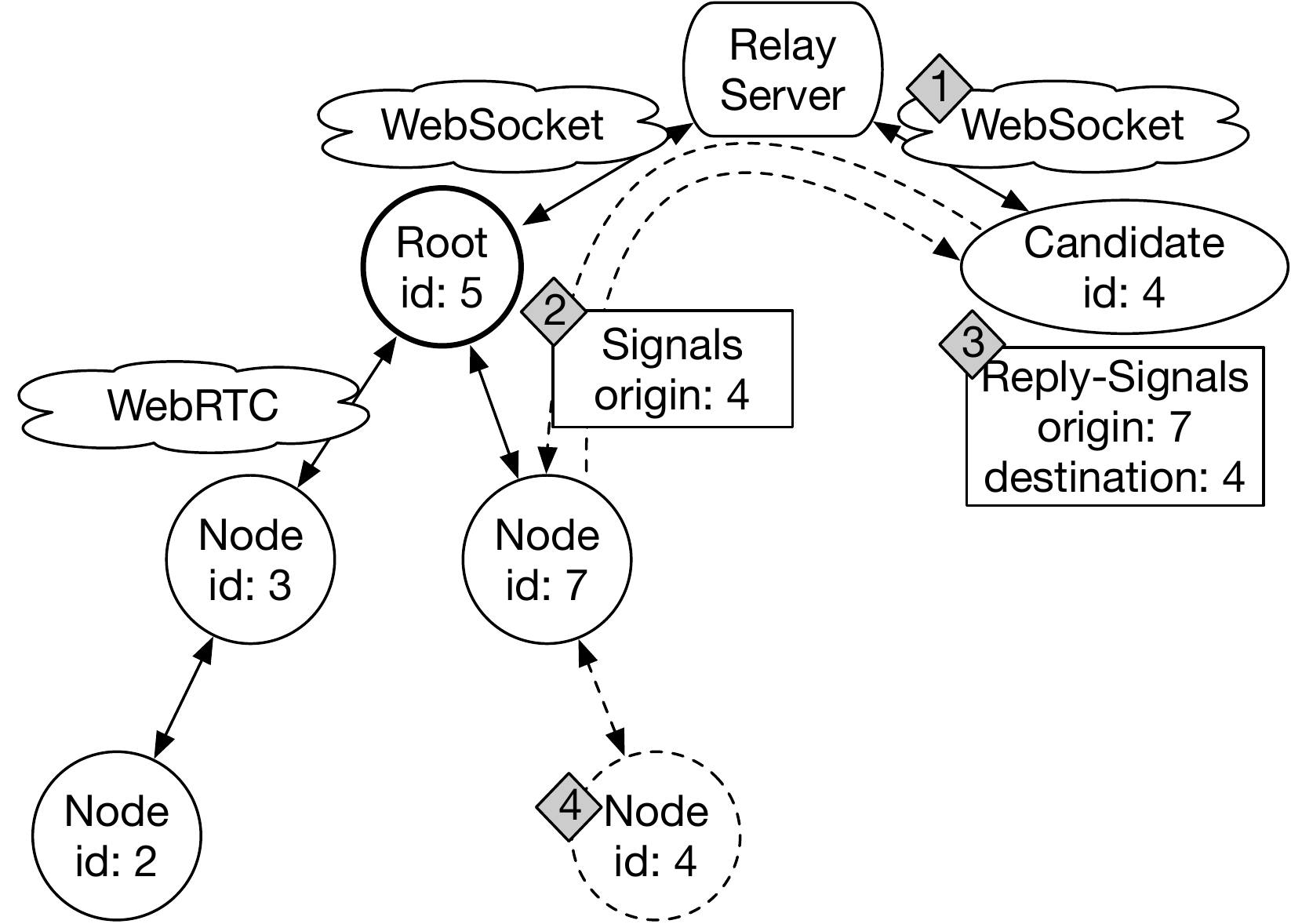}
\caption{\label{Figure:FatTreeDesign}  Genet's WebRTC bootstrap with the joining sequence marked with numbered diamonds.}
\end{figure}

Our solution solves both problems while requiring only information available locally in each node. Each node maintains a list of at most \textit{ChildrenLimit} children, a deployment parameter with a default of 10. Children are added in that list in the order in which they connect and keep the same index until they either disconnect or crash. As illustrated in Figure~\ref{Figure:FatTreeDesign}, when a new participant joins the tree, the \textit{candidate} first opens a WebSocket channel to the Relay Server and creates a random identifier \textit{id} (Step 1). It then sends multiple join requests that each contain its identifier (\textit{origin}) and one of the WebRTC ICE signals to the Root (Master) node (Step 2). From there, each node has two choices. In the first case, if it has less children than \textit{ChildrenLimit}, it assigns the candidate to one of its children and attempts to open a WebRTC connection using the candidate's signals. During the opening, it will generate signals of its own that are sent as replies to the candidate through the Relay Server (Step 3). Signals are exchanged by both parties until a direct WebRTC connection is established, after which the WebSocket connection of the candidate is terminated (Step 4). In the second case (not illustrated), the node delegates the requests to one of its children. If the WebRTC connection to the child is not yet open, the requests are held until the connection is established and then forwarded.

Each node makes routing decisions for delegation by taking the \textit{origin} identifier, xored with the node's identifier \textit{id}, the hash of the result is taken, and then the numerical index of the child in the children list is computed by taking the modulo \textit{ChildrenLimit}: 
\begin{equation*}
	childIndex = hash(originId \wedge nodeId) \% ChildrenLimit
\end{equation*}

The xor of the \textit{originId} and \textit{nodeId} is not strictly necessary, a concatenation of the bits of both identifiers could work also. The advantage of the xor function is to provide a result with the same number of bits as the identifiers, which may be useful when all operations need to be performed in fixed-width registers. 

This routing scheme has three interesting properties. First, routing is \textit{deterministic}: requests from the same origin are routed to the same child at every step of the tree. Second, the choice of a good hash function ensures \textit{probabilistic balancing} of newer connections between the children. Third, by using only information locally available in each node, the routing decisions are \textit{quick} to make because they don't need global information about the tree, which enables a \textit{quick scaling} of the tree on startup.

In some cases, nodes may fail and suddenly disconnect during execution. In those cases, their children, once they have detected the failure, will in turn disconnect their own children (if they have some) and all disconnected nodes will try to reconnect to the root. In other cases, the WebRTC connection may fail to successfully open. Then, the parent node will remove the potential candidate after a configurable timeout, with a default value of 60 seconds.

When deploying the scheme on a local network, it is possible to combine in the same process the Root node and the Relay server. On a wide-area network however, it is important that the Relay Server has a publicly-facing IP address to enable direct WebSocket connections.  

Our implementation performs a routing optimization to accelerate the exchange of messages: to reply to signals, a node opens a direct WebSocket connection to the Relay server. Then if a candidate receives the first reply-signal before having submitted all its own signals, the candidate will use the origin of the reply as a destination for all subsequent signals. This optimization therefore skips some of the routing steps for the late signals. It is however not necessary, another variation that minimizes the number of WebSocket connections to the Relay Server by routing all replies through the Root would also work. We have made our JavaScript implementations of both the Genet algorithm~\cite{webrtc-tree-overlay} and the relay server~\cite{webrtc-bootstrap} available as reusable libraries for Node.js and the browser.

\section{Application to Personal Volunteer Computing}
\label{Section:Application}

We implemented a scalable version of Pando~\cite{lavoie2018pando}, a tool that leverages personal devices' browsers for executing computations in parallel, based on our JavaScript implementation~\cite{webrtc-tree-overlay} of the Genet fat-tree overlay. When a new browser window, executing on the device, successfully connects, it first joins as  a leaf in the fat-tree and computes results, therefore acting as a \textit{processor}. When additional browser windows join beyond the $ChildrenLimit$ of the root, the extras connect to the current leaves. The leaves then stop computing and instead start relaying data and results, becoming \textit{coordinators}. The process repeats at every level of the tree with new devices joining. We have successfully tested a thousand participants (Section~\ref{Section:Evaluation}) but the design should allow potentially millions of devices to connect in a single overlay, the limiting factors being the bandwidth available on the root node and the number of concurrent connections supported by the Relay Server, which determine the joining rate.

\begin{figure}[htbp]
\includegraphics[width=0.48\textwidth]{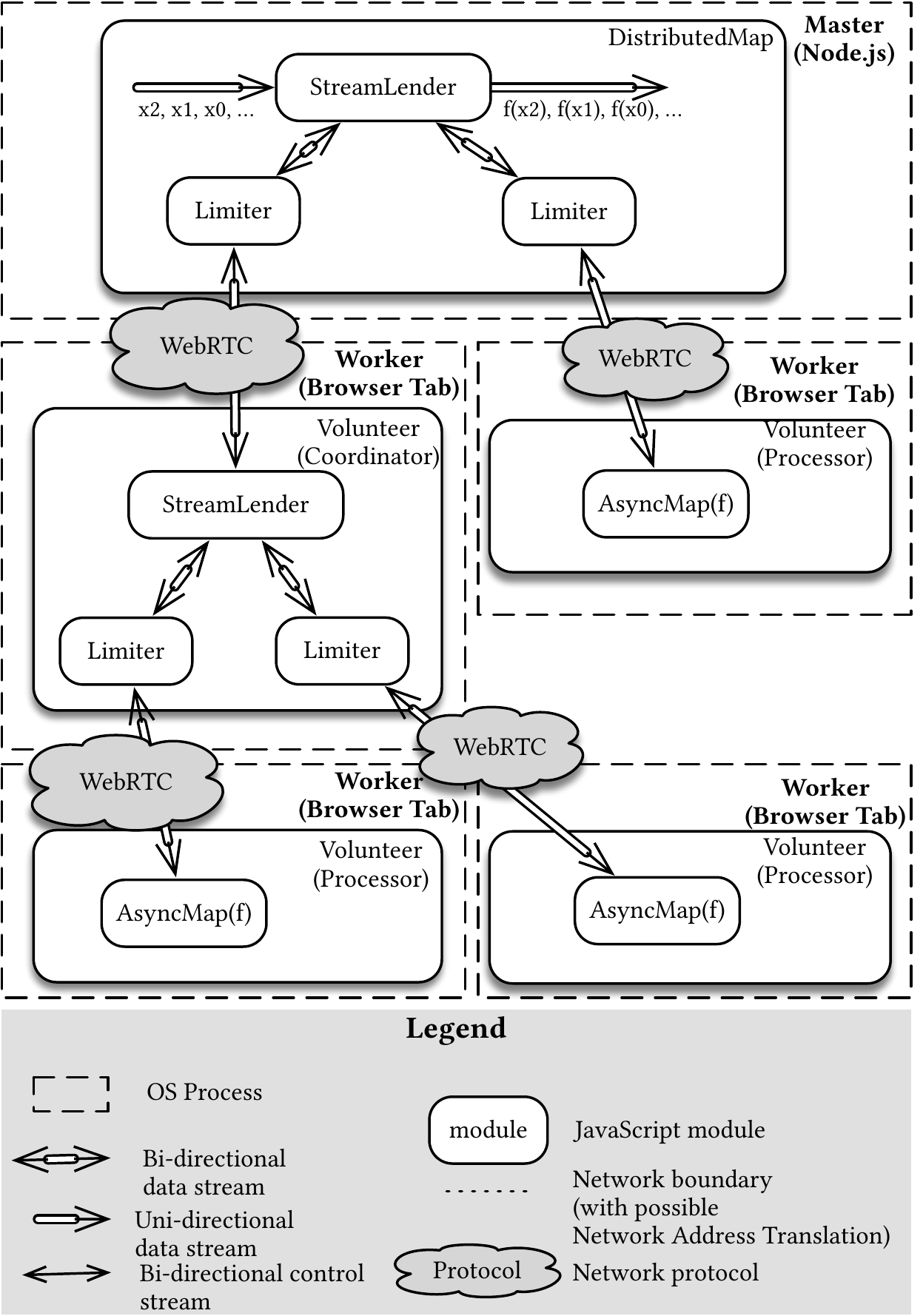}
\caption{\label{Figure:ScalablePando}  Scalable Pando, based on the Genet's fat-tree overlay.}
\end{figure}

The implementation of Pando using the Genet overlay follows a recursive structure. Fundamentally, Pando implements a \textit{streaming map} abstraction: it applies the same function on all inputs it receives from a stream and outputs the results in order. The original implementation of Pando uses the \textit{StreamLender} abstraction to coordinate the distribution of values between a dynamic number of children. To handle potential failures, StreamLender keeps values in memory until a result has been provided. In case of a child's failure, StreamLender will automatically re-submit the memorized values to remaining children. Our scalable implementation re-uses the StreamLender abstraction on intermediary nodes of the fat-tree, as illustrated in Figure~\ref{Figure:ScalablePando}, enabling failures to be handled in the parent of a failing node. Intermediary nodes may also fail. In that case, the parent node will handle the failure by re-submitting the values in a different sub-tree. 

As in the original implementation, 
The StreamLender abstraction and the streaming model used by Pando, are demand-driven and will provide values as quickly as they are asked, enabling faster processors to process more values. However, the WebRTC channel library we use eagerly reads all available values, regardless of the speed at which they are processed on the receiving side. We therefore regulate the flow by using the Limiter module: it lets a limited number of values flow through, after which newer values are sent only after results have been returned. The limit is dynamically adjusted by periodic reports on the number of children in the sub-tree provided by our implementation to adjust the flow to a growing or shrinking fat-tree.

Using the previous design, the fat-tree overlay enables larger total throughput for Pando while providing a quick speed of deployment.

\section{Evaluation}
\label{Section:Evaluation}

In the next sections, we evaluate the behaviour of the design, both in simulation over a large number of experiments, and in real-world deployments, over a smaller number of experiments. We also mesure the benefits provided by the fat-tree overlay when deployed as part of a real throughput-oriented application in personal volunteer computing.

\subsection{Depth with Probabilistic Balancing}

We first study the impact of choosing a \textit{probabilistic balancing} scheme on the depth of the fat-tree under various levels of failure, because the depth has a direct impact on the latency of communication between the root and the leaf nodes.

\subsubsection{How deep is the fat-tree?} $N$ nodes in a perfectly balanced tree are at depth less or equal to $\lceil log(N) \rceil$. Because they are distributed randomly in our fat-tree, a certain percentage of nodes are deeper. To quantify the percentage of nodes that may be affected, we simulated the construction of the tree with nodes with random identifiers that join one after the other, assuming all nodes do not crash. We then counted the number of nodes in the extra levels, and repeated the experiment a thousand times. 

\begin{figure}[htbp]
\includegraphics[width=0.48\textwidth]{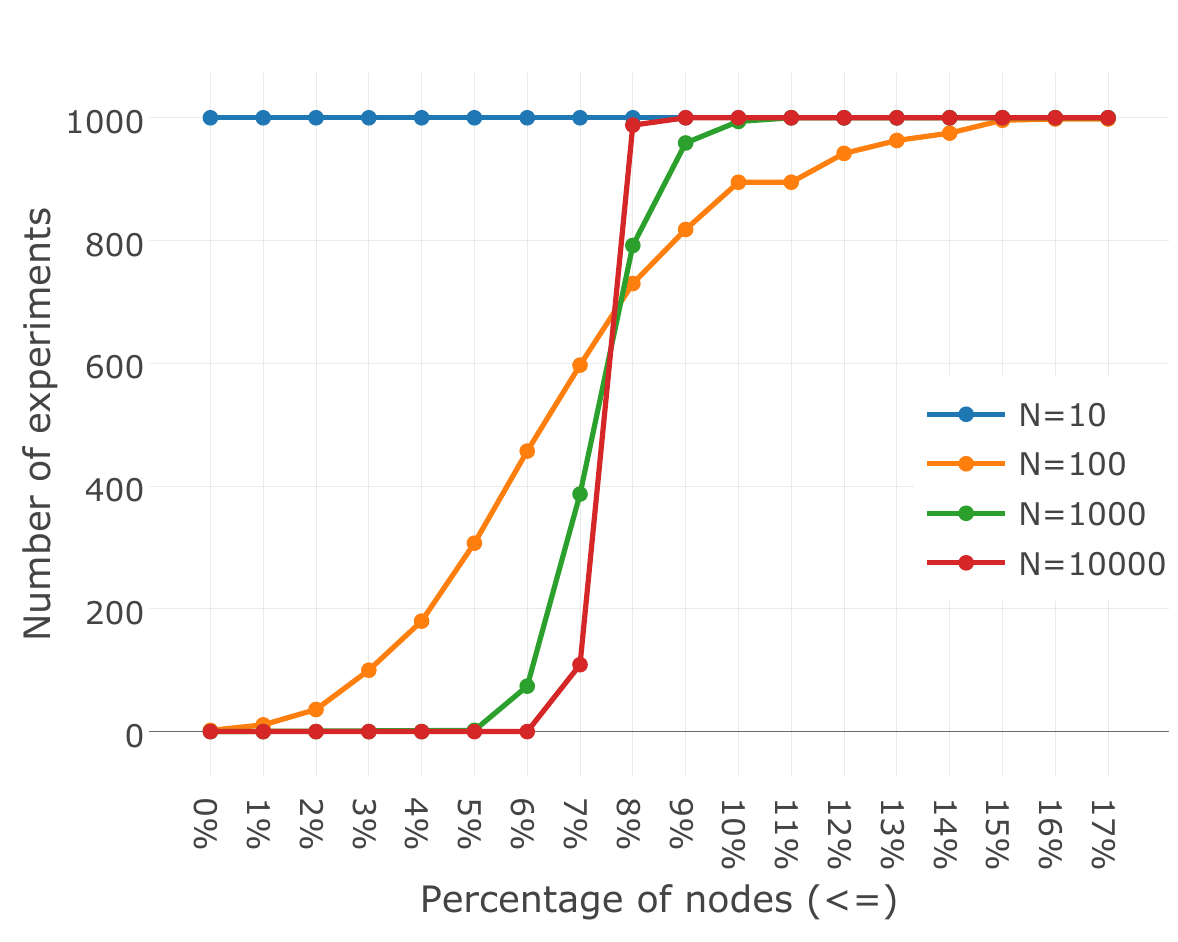}
\caption{\label{Figure:DepthNoFailures} Number of experiments with X\% or less of nodes at depth $log(N) + 1$ over 1000 repetitions and no failures.}
\end{figure}

Over a thousand experiments, we observed no nodes two levels deeper, which while possible in theory is in practice extremely unlikely. The proportion of nodes at depth $log(N) + 1$ varied between experiments. The results are shown in Figure~\ref{Figure:DepthNoFailures}, as a cumulative distribution function for various sizes of trees, to provide both intuitions about the average behaviour and the maximum cases. Our results show that in a majority of experiments ($\geq$700), 8\% or less nodes are on the extra level of the tree, regardless of the number of nodes in the tree. They also show that in all cases, 17\% or less of nodes were in the extra level. Moreover, the larger the tree, the closer all experiments get to around 7.5\% of nodes in the extra level. Therefore, in all experiments,  $\approx83\%$ of nodes are located no deeper than they would have been if the tree had been fully balanced.

\subsubsection{Do failures make it deeper?} In practice, a certain number of nodes \textit{will fail} and force their children to reconnect. To quantify the impact, we construct a tree as in the previous experiment but then disconnect a certain percentage of nodes, then let all nodes reconnect through the root. We then count the number of nodes at deeper levels than $log(N)$. We performed a thousand experiments for trees of size 10, 100, 1000, and 10000 under various probabilities of failure ($F$) from 0 to 1.

\begin{figure}[htbp]
\includegraphics[width=0.48\textwidth]{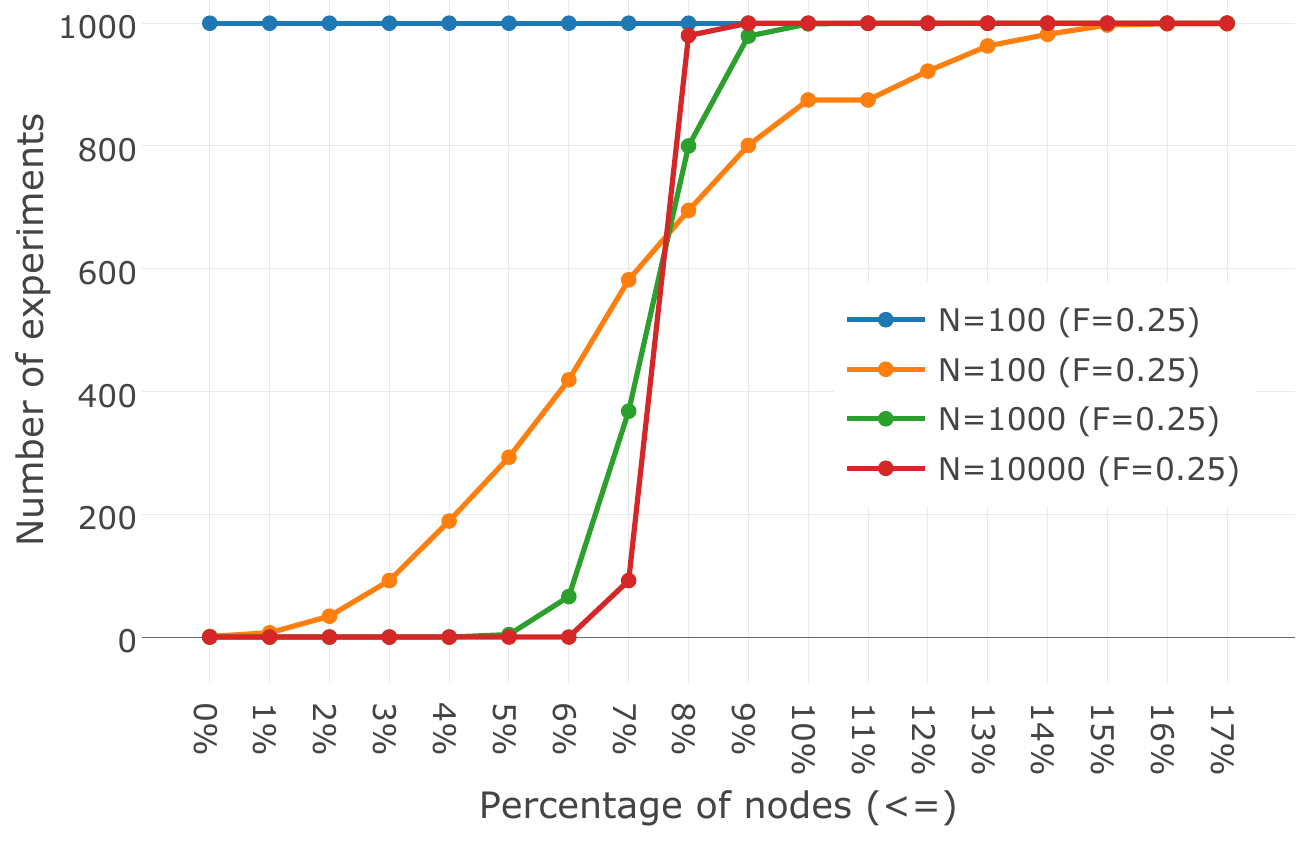}
\caption{\label{Figure:DepthDistributionFailure}  Number of experiments with X\% or less of nodes at depth $log(N) + 1$ over 1000 repetitions and a failure probability of 0.25 for nodes to fail.}
\end{figure}

Over a thousand experiments, we observed no nodes at depth $log(N) + 2$. In all cases, the failures did not affect the percentage of nodes at depth $log(N) + 1$. Results with a 25\% probability of failure are shown in Figure~\ref{Figure:DepthDistributionFailure}; the results are the same for other levels of probability. Failures therefore do not change the distribution of nodes through the tree.

Our probabilistic balancing scheme therefore achieves equivalent depth as a deterministic algorithm for at least $83\%$ of nodes, in the presence of failures or not. For larger trees, of a thousand nodes or more, this percentage increases to $92.5\%$: \textit{scale therefore increases the effectiveness of the scheme} (up to a limit).

\subsection{Bootstrap Latency When Scaling}
\label{Section:BootstrapLatency}

How quickly does the Genet fat-tree scale in practice? We first measure the latency in establishing a WebRTC connection as a baseline and then measure the added overhead of our scheme to connect all nodes, as a function of the size of the fat-tree, with fat-trees of size 10 to a 1000 nodes. We performed these measurements on Grid5000~\cite{grid5000} because it is representative of deployments on a local area network, such as those of a university or a large organization, the infrastructure was accessible to us, it facilitates the replication of experiments, and it can easily scale the number of participating nodes.

For our experiments, we used Grid5000 nodes from the Grenoble site, that has two models of nodes. The first model (\texttt{dahu}) are based on the Dell PowerEdge C6420 that uses Intel Xeon Gold 6130 CPUs (Skylake, 2.10GHz, 2 CPUs/node, 16 cores/CPU), have 192GB of memory and are connected by 10 and 100 Gbps network links. The second model (\texttt{yeti}) is based on the the Dell PowerEdge R940 with also with Intel Xeon Gold 6130 CPUs (Skylake, 2.10GHz, 4 CPUs/node, 16 cores/CPU), have 768GB of memory and are also connected by 10 and 100 GBps network links. The exact distribution of nodes for experiments is chosen randomly between the two models, based on availability (because other experiments concurrently run at the same time on other machines), in our case it was almost always the \texttt{dahu} nodes that were used, with an occasional \texttt{yeti} node in the mix. All our throughput experiments were also made with these nodes.

\subsubsection{How long does it take to establish a single WebRTC connection?}

While individual nodes on Grid5000 enjoy a sub-millisecond latency, the ping between nodes is typically between 0.1 and 0.2 ms, establishing a WebRTC connection is significantly slower. As explained in Section~\ref{Section:FatTreeOverlay}, each participant in a connection first starts listing potential connection end-points that can enable Network-Address Translation, also contacting STUN servers in the process.  For example, the Google STUN server we use (\texttt{stun.l.google.com}) has an average ping latency of 35ms. The endpoints need to be exchanged between participants through a relay and finally, multiple connections are tried from both sides until one is found to work. Between nodes on a local network, the connection can be established earlier because some of the endpoints will use the local IP address and therefore a connection on the local area network can be established before the other endpoints discovered by the STUN server are received. Nonetheless, the messages exchanged with the relay server significantly adds to the delay.

\begin{figure}[htbp]
\includegraphics[width=0.5\textwidth]{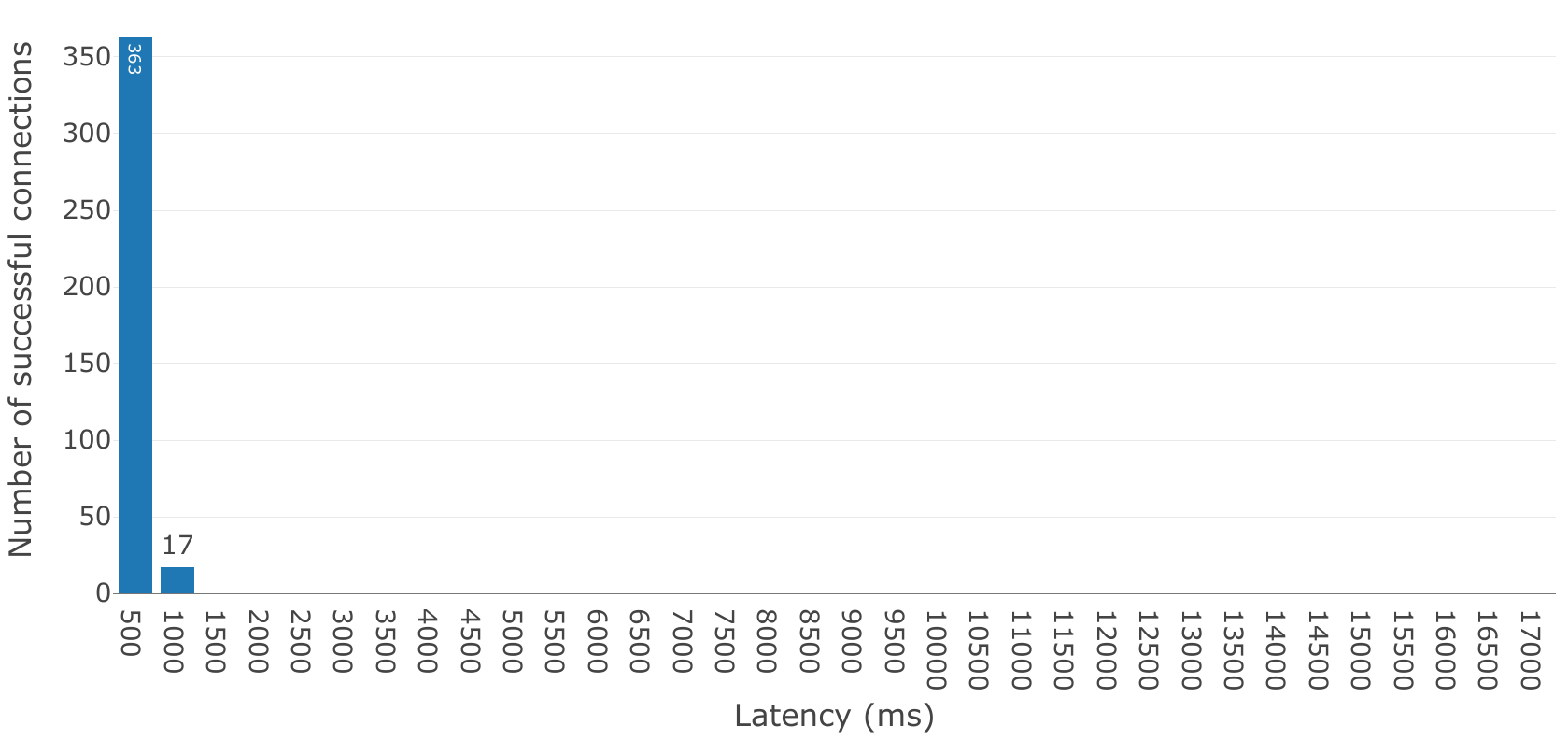}
\caption{\label{Figure:WebRTCConnectionLatencyGrid5000Local} WebRTC connection latency distribution over 380 successful connections between nodes on the Grid5000's Grenoble site using a local server.}
\end{figure}

In our tests, the relay server was running on the local network and connected 20 browser windows on 10 Grid5000 nodes forming a fully connected clique, each window opening a connection to every other window. All connection attempts succeeded. For all connections, we measured the latency between creating the connection and a confirmation message sent through the connection, which corresponds to the time it takes before data starts flowing through the connection. We used \texttt{webrtc-connection-testing}~\cite{webrtc-connection-testing} version 4.0.0, an open source tool we built for this task. The results are shown in Figure~\ref{Figure:WebRTCConnectionLatencyGrid5000Local}. We observed connection latencies less than 1000ms, with 95.5\% of connections taking less than 500ms. Of the 363 connections that took less than 500ms, 41 took less than 100ms, 112 took between 100ms and 200ms, 132 took between 200ms and 300ms, and 78 took between 400ms and 500ms (not shown on the figure).  

Therefore, even if Grid5000 nodes have sub-millisecond ping latencies, establishing a WebRTC connection is slow and typically takes three to four orders of magnitude longer, up to 1000ms. As the implementation of WebRTC is part of the implementation of the browser, any overlay design executing in JavaScript is subjected to this constraint and will be fully deployed at the speed at which the slowest connections are established.

\subsubsection{How long does it take to connect all nodes in a fat-tree?}
\label{Section:ConnectionLatency}

Fully deploying a fat-tree, in which nodes are organized in multiple layers, shall therefore take at least time that is proportional to the depth of the tree and the time it takes to establish the slowest connections. In this section and the next, we show it is the case in practice with a full implementation in a working tool that we use for actual applications.

We used Pando version 0.17.9~\cite{pando-repository} for our tests, which implements the design of Section~\ref{Section:Application}. The fat-tree is used to distribute inputs to processors and retrieve back results. We used a fat-tree of degree 10, which means that each node has at most 10 children and each layer of the tree will have a multiple of 10 participants in it. We used a test application that waits for 1 second then returns the square of the input value for a number of reasons. This removes the impact of potential differences of CPUs speeds, making it easier to determine when all processors are connected and producing results because the overall throughput, in values per seconds, is equal to the number of participating processors (on the leaves of the fat-tree). In turn, this means our measurements really represent the \textit{coordination overhead} of the entire system. And finally, the time it takes to reach the full throughput represents the bootstrapping latency.

We deployed the fat-tree on 10 different Grid5000 nodes on the Grenoble's site by progressively increasing the number of browser windows executing on each node, with 1, 5, 10, 25, 50, and 100 windows. We repeated each experiment five times. Each node was connected with a one second delay after the previous, e.g. the first node opens its browser windows after a 1-second delay, the second with a 2-seconds delay, etc. up to 10 seconds for the last node. While opening a large number of browser windows at the same time (each executing in their own operating system process) from the same node worked fine, launching browsers at the same time from multiple nodes led to connection errors, which prompted the addition of an artificial delay between Grid5000 nodes. The rate of connection for 10 browser windows is therefore 1 browser window per second for 10 seconds, while the rate of connection for 1000 browser windows is 100 browser windows per second for 10 seconds. 

\begin{figure}[htbp]
\includegraphics[width=0.5\textwidth]{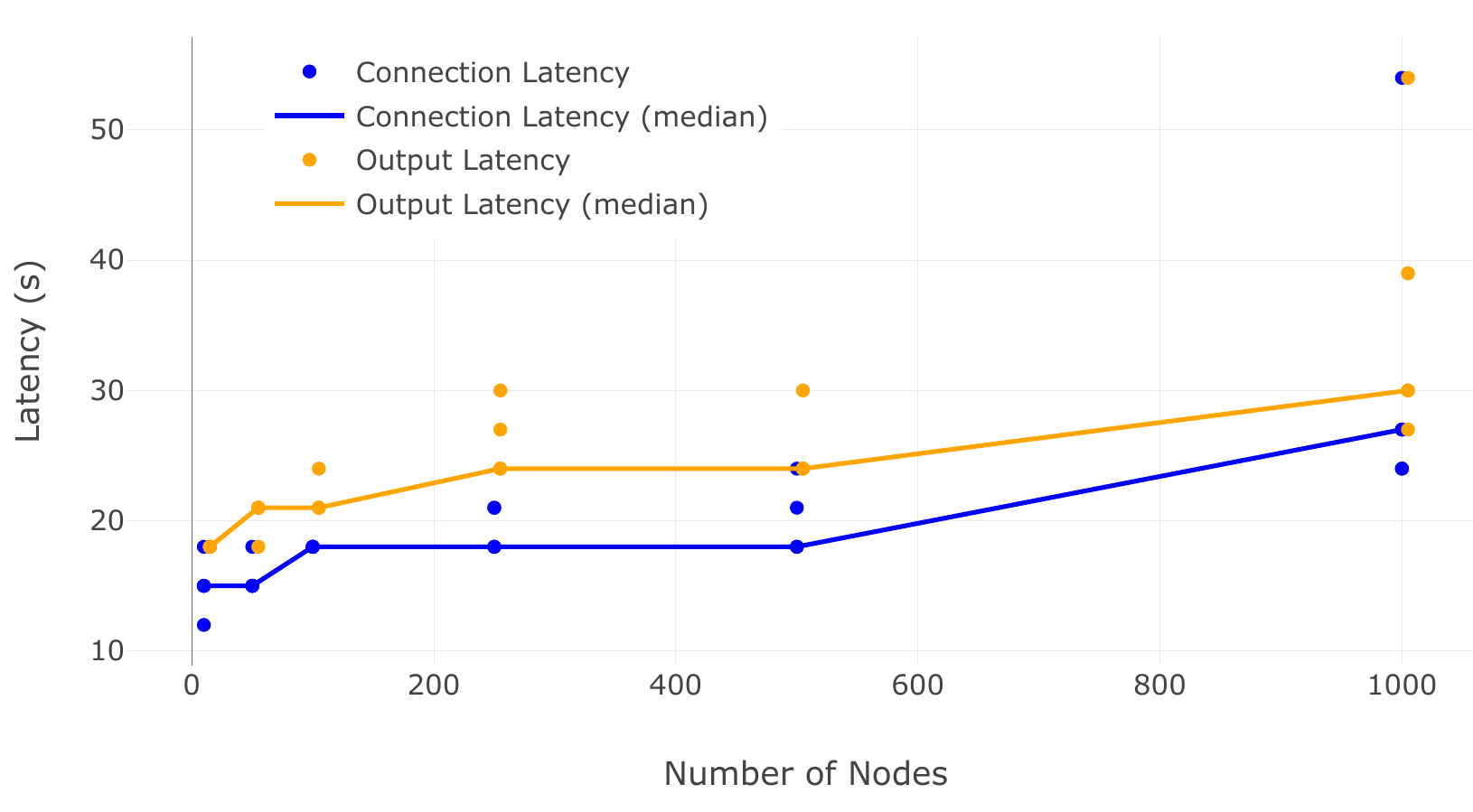}
\caption{\label{Figure:ConnectionLatency} Latency to connect all participants in the fat-tree  on the Grid5000's Grenoble site over 5 experiments.}
\end{figure}

As the fat-tree is deploying, periodic status updates are sent from the leaves to the root node to report on the current state of the fat-tree. We used an interval of 3 seconds between reports, therefore the state of the fat-tree may be known at the earliest 3 seconds after having changed. This means the latency that we report in the next figures represent an upper bound on the actual latency to connect the nodes. We measured the time it takes until all browser windows were counted as children in the tree. We then also measured the throughput of squared values at the output of Pando, also by sampling at intervals of 3 seconds. We measured the time it takes until the throughput corresponds to the number of leaves in the fat-tree, as reported by the previous reporting mechanism. The throughput measurements are independent from the reporting strategy, and the latency measured really represents the time observed by a user of Pando until full throughput is achieved. Both results are shown in Figure~\ref{Figure:ConnectionLatency}.

For 10 browser windows, it typically takes about 15 seconds to connect them all in the first layer of the fat-tree. This is about 5 seconds longer than the 10 seconds required to open all browser windows; the additional delay can be explained by a 1-2 second delay before Pando's server is ready after start-up, the maximum latency of 1000ms to establish the slowest connections, as reported in the previous experiment, and the reporting interval, i.e. we learned of that last connection 5 reports after Pando was started. The maximum throughput is first observed one sample later, at 18 seconds, as shown by the 'Output Latency' curve. As the size of the tree increases, so does its depth and therefore the latency to fully connect the fat-tree. At 100, 250, and 500 browser windows its takes about 18 seconds to fully connect the children, while at 1000 browser windows it takes 24 seconds. The latency to reach maximum throughput follows accordingly by one or two samples, 21 seconds for 50 and 100 browser windows, 24 seconds for 250 and 500 browser windows, and 28 seconds for 1000 browser windows. The variation between experiments also grows larger as the size of the tree grows, we measured a latency of up to 54 seconds both for connecting children and reaching maximum throughput in a single experiment. We can therefore conclude that it typically takes 30 seconds to connect all nodes in our WebRTC fat-tree and sometimes up to a minute on the Grid5000 testbed.

\subsection{Throughput Ramp-up in the Collatz Application}

Now that we have established a typical latency of 30 seconds to reach maximum throughput, does the fat-tree, when used with Pando, behave in the same way when the leaf nodes are actually performing computations, as described in Section~\ref{Section:Application}? We answer that question by taking one representative application of volunteer computing, the Collatz Conjecture~\cite{boinc-collatz}, which has also been implemented~\cite{boincprojects} using the BOINC~\cite{anderson2004boinc} infrastructure. This is essentially a number-crunching application, with really small inputs requiring a significant amount of computations, most of the time being spent manipulating big integers that do not fit into registers. We implemented the application in JavaScript with an off-the-shelf Big Number library. For the purpose of measuring the scaling behaviour, the single core performance is not critical, a faster implementation shall increase our throughput measurements by a constant factor, which would be obtained by better using the CPU, while not affecting much the scaling behaviour, which is instead due to the coordination performed by the fat-tree.

\begin{figure}[htbp]
\includegraphics[width=0.5\textwidth]{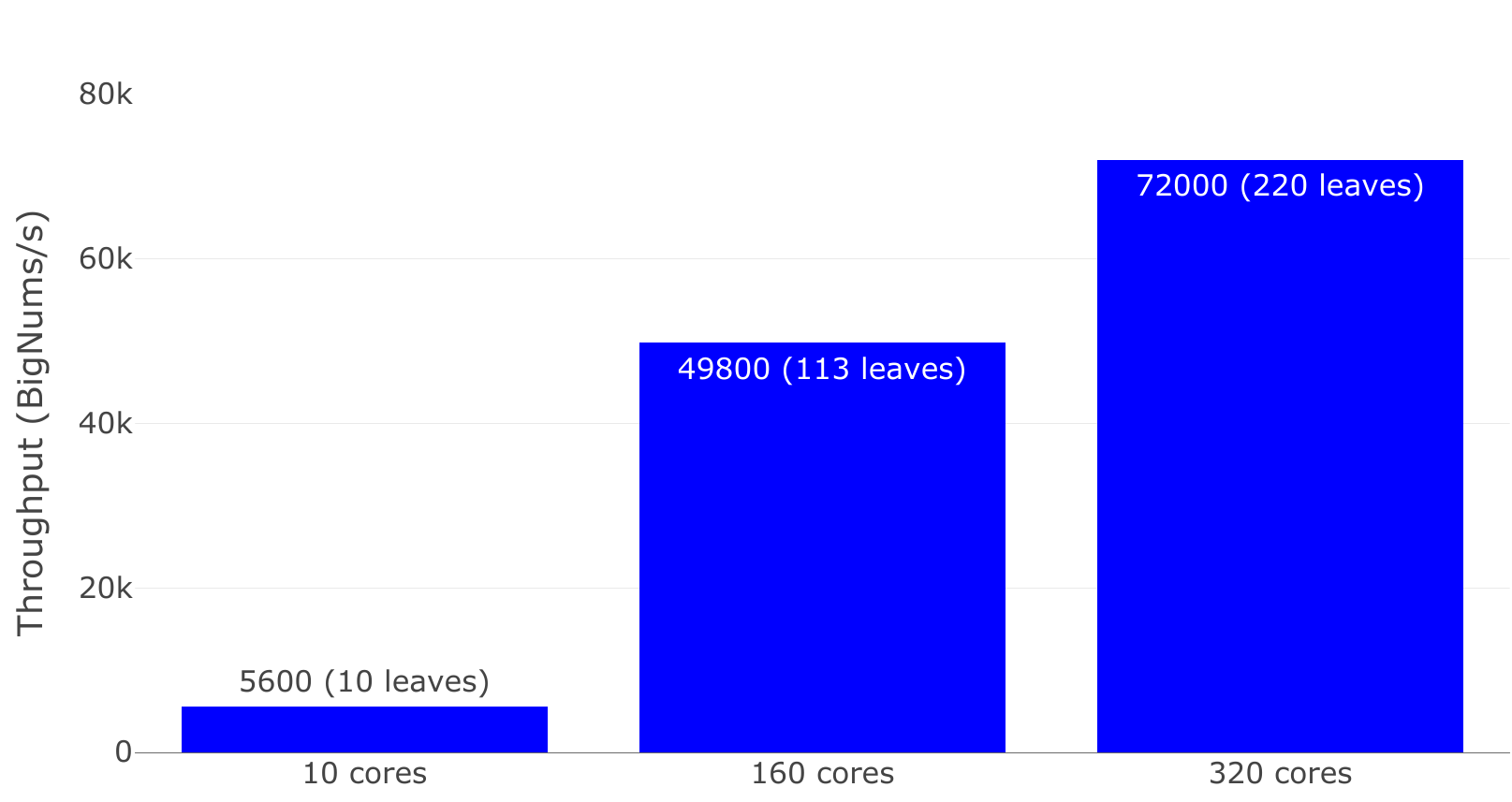}
\caption{\label{Figure:CollatzAverageThroughput} Average throughput on the Collatz application on Grid5000 as the number of cores used increases.}
\end{figure}

Studying the throughput scaling behaviour on actual applications is complicated by the fact that all tasks do not take the same amount of time. The throughput at the output of Pando can vary both because nodes join or because tasks are temporarily faster or slower. Moreover, as our fat-tree design probabilistically balances the tree, the actual number of leaves that are processing inputs varies between experiments. It is therefore harder to determine when all connected nodes have started contributing to computations. We therefore first measured the average throughput with a given number of nodes, and let the deployment compute for at least three minutes. We then took the average throughput measured after all the nodes were connected and counted the number of participating processors (leaves). We measured for 10 Grid5000 nodes, with 1, 16, and 32 browser windows per node, the last being the maximum number of cores available on the machines of the Grenoble site. The results are shown in Figure~\ref{Figure:CollatzAverageThroughput}.

As the number of browser windows increases, so does the average throughput, showing a clear benefit to scaling the number of participating cores. However, the results are not quite linear. This is actually not due to the fat-tree design but contention for CPU resources on the same machine. We did a quick second experiment with 10 browser windows on a single machine and we obtained $\approx 480 \frac{BigNums}{(s*node)}$ rather than the $\approx 560 \frac{BigNums}{(s*node)}$ we obtained with 10 browser windows on 10 nodes. 

\begin{figure}[htbp]
\includegraphics[width=0.5\textwidth]{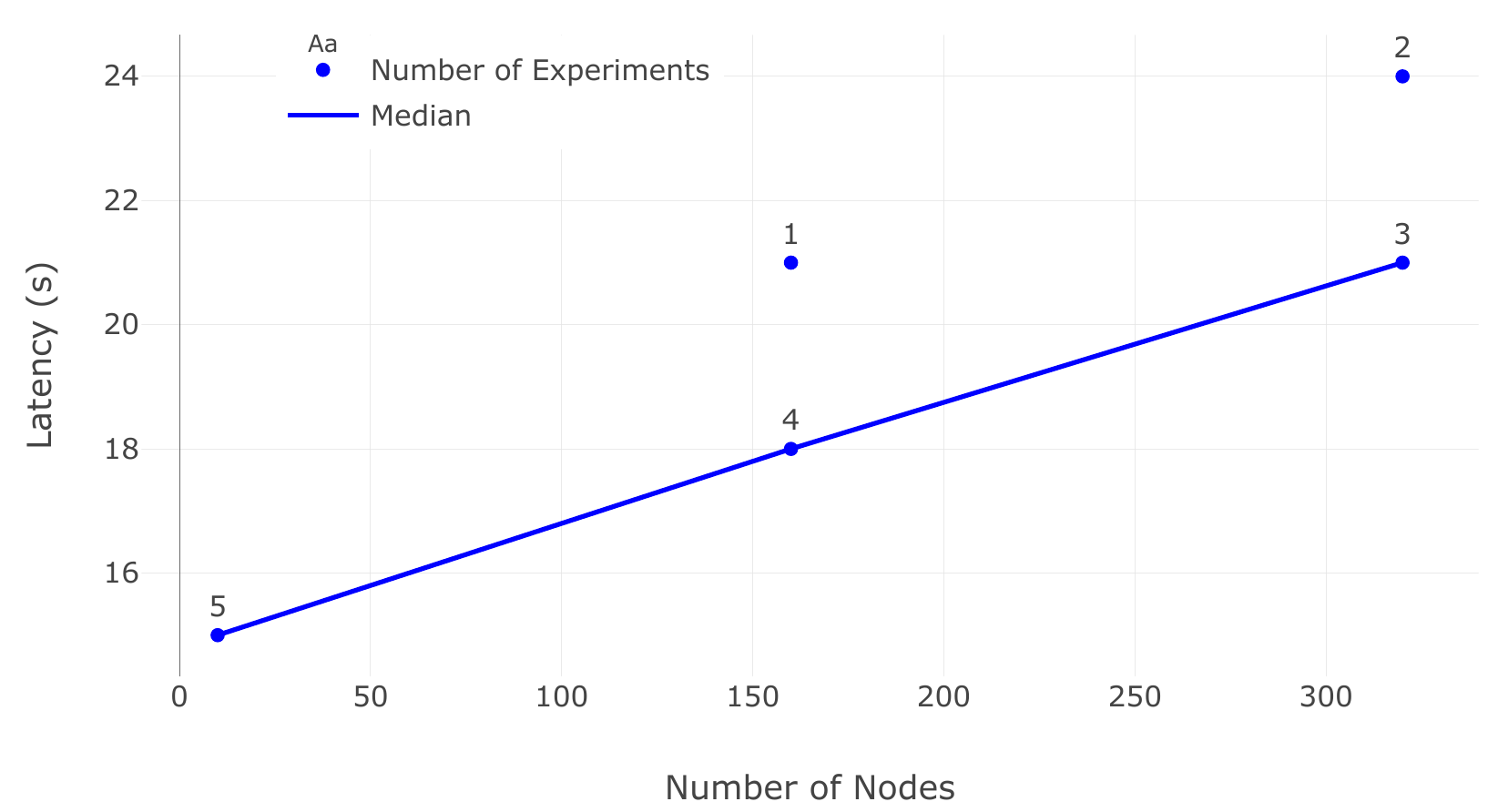}
\caption{\label{Figure:CollatzBootstrapLatency} Latency to reach maximum throughput on the Collatz application on Grid5000 as the number of cores increases.}
\end{figure}

We then used the previous average throughput as a target to determine the time it takes before all cores are actually contributing results, when deployed with the fat-tree overlay. We therefore measured the time it takes until the output throughput reaches the average throughput measured previously, adjusted for the actual number of participating processors (leaf nodes in the fat-tree). We used the same methodology as in Section~\ref{Section:ConnectionLatency}. The results are shown in Figure~\ref{Figure:CollatzBootstrapLatency}.

The results are consistent with the previous results, even slightly better probably because of the uncertainty added by the 3 seconds sampling interval. In this case again, reaching maximum throughput typically takes 15 seconds with 10 nodes ($\approx$ 10 cores), 18 seconds with 160 nodes ($\approx$  110 cores), and 21 seconds with 320 nodes ($\approx$  220 cores).

\subsection{WebRTC Connection Probability and Establishment Latency on the Internet}

The previous results show that our WebRTC fat-tree design is effective in quickly deploying a large number of nodes on a local area network and provided a methodology for systematically studying their performance for volunteer computing applications, both of which had never been done before. In this section we provide some additional intuitions about how a deployment that targets the Internet should be adapted and show how the tools we built for the previous experiments can be used in that setting to motivate future works.

The previous results already show that the latency in establishing WebRTC connections is a significant factor in the overall latency of deploying a fat-tree, because even on a fast local network, a connection can take up to 1000 ms to be established. We tested two additional settings, one in which the relay server for exchanging the connection endpoints is located outside the local network and a second in which the participants are distributed across the planet.

\begin{figure}[htbp]
\includegraphics[width=0.5\textwidth]{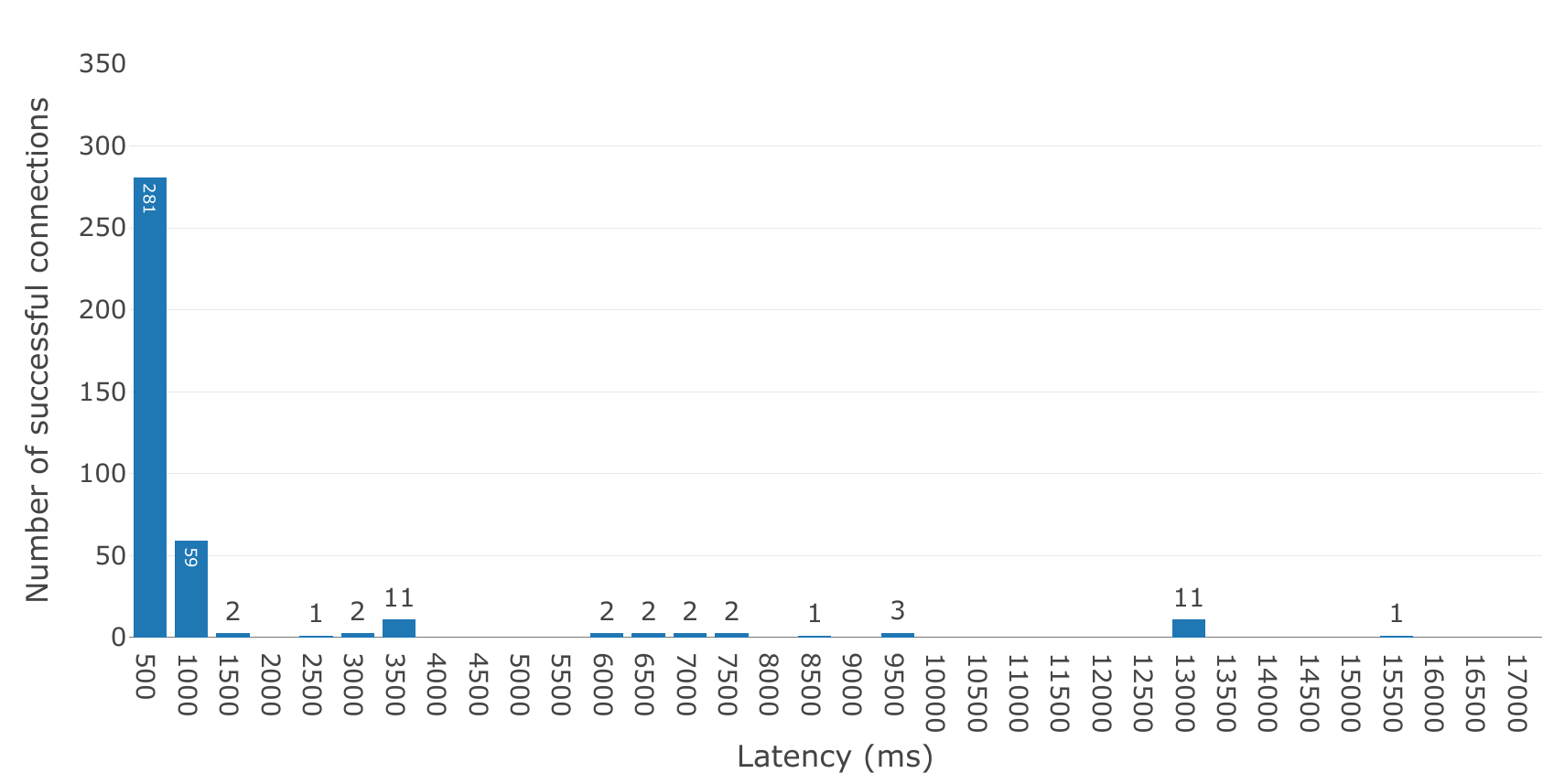}
\caption{\label{Figure:WebRTCConnectionLatencyGrid5000Remote} WebRTC connection latency distribution over 380 successful connections between nodes on the Grid5000's Grenoble site using a remote server.}
\end{figure}

We show the results of the first experiment in Figure~\ref{Figure:WebRTCConnectionLatencyGrid5000Remote} when establishing connections between browser windows executing on Grid5000, but relying on a remote server located in Paris, France\footnote{Running on Amazon Cloud.}, for relaying signalling messages. The ping latency from Grenoble to that server takes 40ms on average and ranges from 13ms to 150ms, about 130-1500 times higher than between nodes within Grid5000. All connections succeeded also in this case. We observed similar results as for the experiment with a local server but with greater variability, with some connections taking between 1000ms and up to 16s to be established. Among the fastest established connections, 16 took less than 100ms, 123 took between 100ms and 200ms, 82 took between 200ms and 300ms, 31 took between 300ms and 400ms, 29 took between 400ms and 500ms, 36 took 500ms to 600ms, and 17 took 600ms to 700ms (not shown in the figure), and together account for 89\% of all latency results. Compared to using a local server, 22.6\% less connections take less than 500ms and almost three times more take between 500ms and 1000ms. Connections therefore have additional latency as well as greater variability, as would be expected from messages routed on the Internet.

\begin{figure}[htbp]
\includegraphics[width=0.48\textwidth]{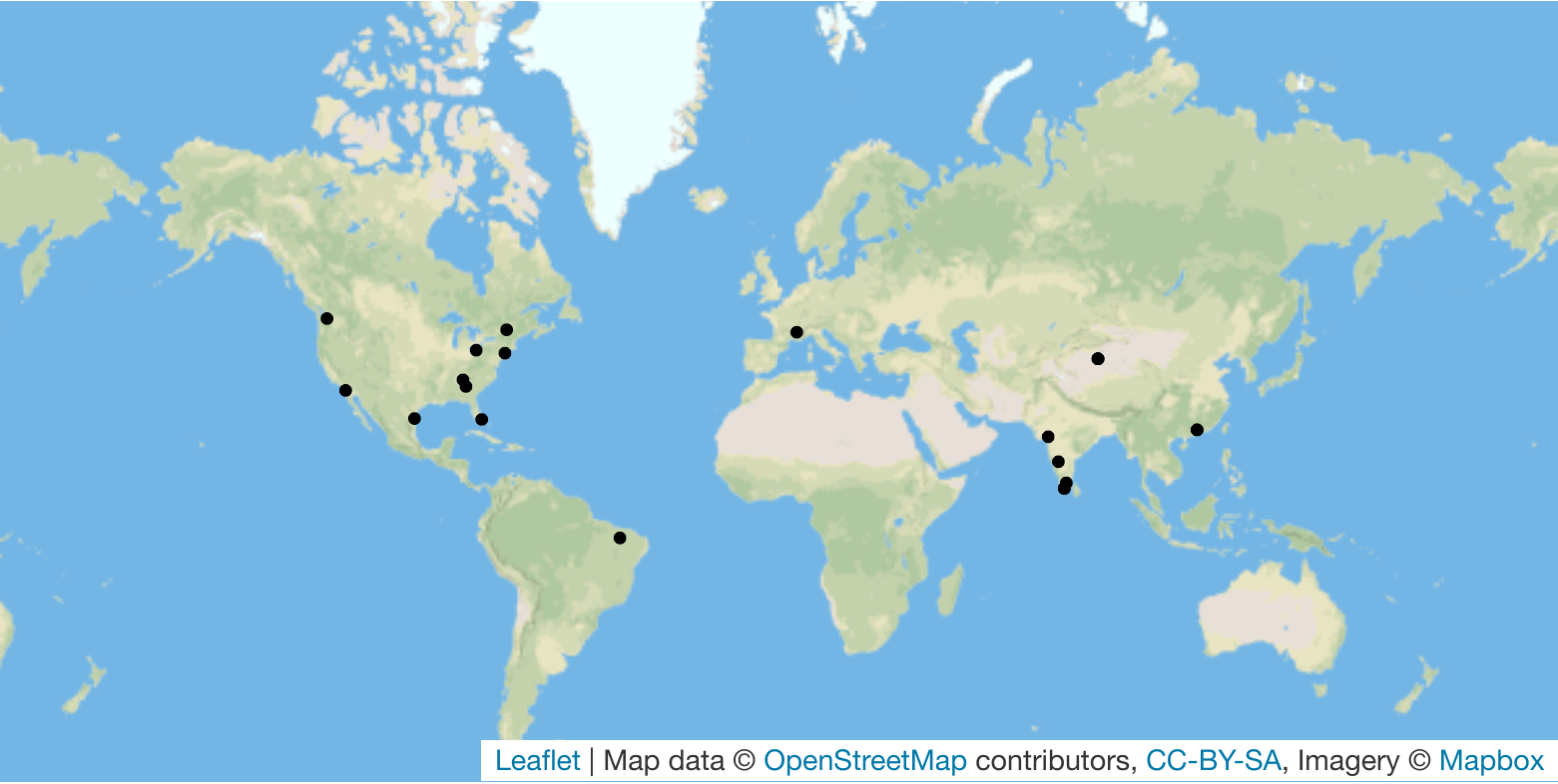}
\caption{\label{Figure:GeographicalDistributionWorkers} Geographical location of participants that accepted to share their location.}
\end{figure}

For the second experiment, we asked 20 participants randomly selected among Mechanical Turk~\cite{mechanical-turk} workers to open a web page that tested their WebRTC connectivity to other participants and the experimenter. Out of the 21 participants, including the experimenter, 17 chose to voluntarily disclose their location using the geolocation API of their browser. The world-wide distribution of participants is shown in Figure~\ref{Figure:GeographicalDistributionWorkers}. Between all participants that were connected to the relay server at the same time, 398 WebRTC connection attempts were made, out of which 194 succeeded, for a success ratio of 48.7\%. This shows, unsurprisingly, that random connections between participants do not always succeed. However, contrary to our initial expectations, almost half of the connections succeeded. 

\begin{figure}[htbp]
\includegraphics[width=0.5\textwidth]{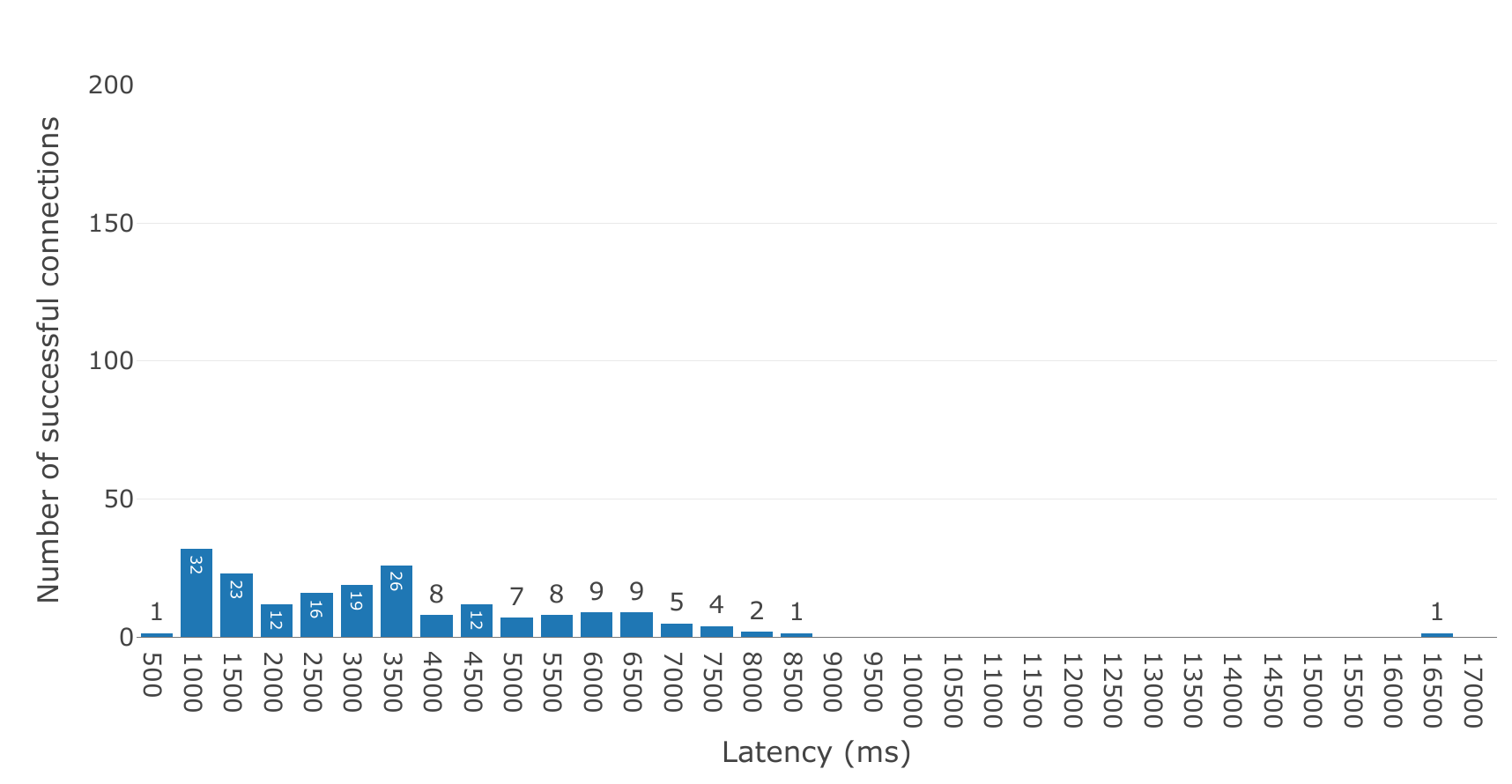}
\caption{\label{Figure:WebRTCConnectionLatency} WebRTC connection latency distribution over 194 successful connections between world-wide participants.}
\end{figure}

The latency in establishing connections, as shown in Figure~\ref{Figure:WebRTCConnectionLatency}, has more variation compared with the local and remote server experiments on Grid5000. Except for one result, all other connections took at least 500ms to be established and most results are well-distributed between 500ms and 8500ms. Compared to Figure~\ref{Figure:WebRTCConnectionLatencyGrid5000Local}  and Figure~\ref{Figure:WebRTCConnectionLatencyGrid5000Remote}, it is therefore more typical for a participant to take several seconds to be connected.\footnote{This experiment used the previous 3.0.2 version of the \texttt{webrtc-connection-testing} tool~\cite{webrtc-connection-testing}, which can introduce an additional connection delay because participants that are already connected receive notifications of newer participants only every 5 seconds. This was fixed in version 4.0.0 (which was used in the previous experiments) to send notifications as soon as participants are connected. 
Because we could not assemble again the same set of participants to repeat the Internet deployment of the original paper submission, we present the original results. While the average latency could possibly be lower, we would expect to still observe an increased variability of the latencies and values of at least a few seconds.}

Supposing our results generalize, which shall be validated in larger settings, this means that choosing a random Internet participant for connection shall lead to a successful connection almost half the time. However, this also highlights the need for mechanisms to tolerate failures of initial connections. One possible solution would be to first test for connectivity before deploying the fat-tree, which unfortunately would lead to a higher connection latency. A second possible solution would be to attempt multiple random connections when a participant joins. In effect, bootstrapping the fat-tree this way would lead to a mesh network, which could then, for example, be made to converge to a more efficient topology if necessary.

\section{Related Work}
\label{Section:RelatedWork}


\textit{Fat-tree topologies} have originally been proposed to provide high-bandwidth communication between nodes in computing clusters while minimizing the cost of switching hardware~\cite{Leiserson85}. Fat-trees derive their name from the increasing bandwidth requirements on edges closer to the root because they relay traffic for all children in the underlying sub-tree. Fat-trees were later also adopted explicitly or implicitly in \textit{overlay networks}, in which nodes connected using Internet protocols are organized in logical networks for efficient communication, to provide, for example, multicast communication~\cite{zhang2012survey,birrer2004fatnemo}.

Extensive work on tree overlays for multicast applications has been done since the 90s~\cite{zhang2012survey}, in which the same data is disseminated from a single source to tens and up to millions of participants. Typical applications of volunteer computing have different data transfer patterns because each participant receives a different sub-set of data. BOINC submits the same computation to a small number of participants (at least three) until a majority agrees~\cite{taufer2005homogeneous}, while the current version of Pando~\cite{lavoie2018pando} does not use redundancy because the code is executed on trusted devices. In addition, in both cases, each participant will return different results to the root.

To the best of our knowledge, we are the first to propose a fat-tree overlay for scaling volunteer computing applications that supports an infinite number of inputs and provides a \textit{decentralized} scheme for allocating nodes in the tree.  ATLAS~\cite{baldeschwieler1996atlas}'s tree of managers organized around work-stealing is perhaps the oldest documented scheme that relies on a tree for scalability but little details about the implementation were provided and the actual implementation was tested with only 8 machines. Javeline++~\cite{neary2000javelin++} relies on a tree structure to implement a \textit{distributed work-stealing} scheduler but the scheme relied on tasks being finite and the position of a new node in the tree is computed from the root. Bayanihan~\cite{sarmenta1998bayanihan} conceived a tree of servers that maps to the underlying network topology when the bandwidth on the link to a single server is insufficient, but to the best of our knowledge the scheme was never implemented. Connection decisions in our scheme do not require global information about the tree, yet they ensure probabilistic balancing and guarantee the routing of multiple connection messages to the same leaf node.

BOINC~\cite{anderson2004boinc} currently supports hundreds of thousands of participants but relies on a dedicated server with sufficient resources and an interaction pattern that is tailored to long running computations. Volunteers obtain the task to perform and transmit the results in two different remote procedure calls. Participant failures are detected with a soft limit on the expected time to completion, which therefore requires an estimate that is application dependent. Our design is tailored to shorter running tasks and instead relies on the heartbeat mechanism provided by WebRTC to detect the failure of a participant. Moreover, by relying on WebRTC to scale up the number of concurrent connections, we can support at least a thousand participants with no investment in dedicated hardware nor renting of hosted resources.

Compared to other published volunteer computing tools, we are the first to have successfully tested with a thousand participants and the first to use WebRTC to connect participants in a fat-tree overlay. Most published volunteer computing tools~\cite{alexandrov1997superweb,baratloo1999charlotte,cappello1997javelin,sarmenta1998bayanihan,duda2012distributed,martinez2015capataz,cushing2013weevilscout,kuhara2014peer,leclerc2016space} were tested with less than a hundred of participants. Some of the most recent have been tested with more than a hundred participants~\cite{merelo2008asynchronous,langhans2013crowdsourcing,meeds2015mlitb} and even up to 400 concurrent participants~\cite{dkebski2013comcutejs}. But the largest internet deployments of custom tools~\cite{merelo2008asynchronous,langhans2013crowdsourcing,meeds2015mlitb} have so far reached a hundred concurrent participants~\cite{meeds2015mlitb}.

WebRTC~\cite{webrtc} has been used in the design of other kinds of overlay networks, including content delivery~\cite{hive.js}, real-time collaboration~\cite{vanderLinde2017legion}, and virtual reality~\cite{Hu:2017}. Kuhara and al.~\cite{kuhara2014peer} have proposed a service to share files for volunteer computing but they tested their system on a single machine. BrowserCloud.js~\cite{dias2018browser}, is the only other distributed computing platform we are aware of that also uses WebRTC as an overlay. Contrary to our design, it is organized around a distributed hash table rather than a tree, and tasks are pushed from the submitting peer to available workers rather than being pulled by workers as they become free. The implementation of browserCloud.js has been tested on 10-25 browsers on a single machine, which provides little information about the speed at which their overlay can scale in deployments on a local network. Spray~\cite{nedelec2018adaptive} is a peer sampling implementation that also uses WebRTC and they also tested on the Grid5000 testbed, with up to 600 hundred concurrent browsers. However, their experiments limit the rate at which participants join to 1 per 5 seconds. It therefore takes 50 minutes for the 600 browsers to join. In a similar setup, our fat-tree overlay deploys on a thousand browsers in 20-55 seconds.

\section{Conclusion and Future Work} 
\label{Section:Conclusion}

We have presented the Genet Fat-Tree overlay, which enables quick scaling by relying on a novel scheme that only requires \textit{local information} to route connection messages. The routing scheme derives the destination of messages from the hash value of the combined identifiers of the message's source and of the node that is holding the message. The scheme provides \textit{deterministic routing} of a sequence of connection messages from a single source and \textit{probabilistic balancing} of newer connections among the leaves, which is especially useful when implemented with WebRTC because opening a new channel requires the exchange of multiple independent \textit{signal} messages between participants. We have shown that the probabilistic balancing of the tree, induced by the routing scheme, puts at least 83\% of nodes at a similar depth than they would have been with a deterministic balancing algorithm, increasing to 92.5\% on trees of a thousand nodes or more. We have also shown that an implementation of the design could connect a thousand browser windows in a local area network in 22-55 seconds and enable the throughput on the Collatz application to increase by two orders of magnitude, coordinating over 220 computing cores out of 320 participating cores. We have finally motivated future work to generalize the design for a world-wide setting, by taking into account the lower connection probability in a wide-area network and the increased connection latency.

The Genet Fat-Tree overlay could be applied to other problems than volunteer computing. The most promising one seems to bootstrap other overlay networks built with WebRTC. It could, for example, implement a peer sampling protocol, such as Spray~\cite{nedelec2018adaptive}, and the initial bootstrap could be made fast by having new nodes join \textit{multiple nodes} in the tree, forming a mesh that could then progressively converge to an efficient topology. The quick scaling ability of the design we have presented is therefore complementary to potential refinements based on existing overlay designs.

\section{Acknowledgements} 
 This material is based upon work supported by the National Science and Engineering Research Council (NSERC) of Canada, Fond de Recherche du Quebec -- Nature et Technologies (FRQNT).  Any opinions, findings, and conclusions or recommendations expressed in this material are those of the author and do not necessarily reflect the views of NSERC or FRQNT.

Experiments presented in this paper were carried out using the Grid'5000 testbed, supported by a scientific interest group hosted by Inria and including CNRS, RENATER and several Universities as well as other organizations (see \url{https://www.grid5000.fr}).

\bibliographystyle{plain}
\bibliography{ErickLavoie}

\end{document}